\documentclass[epj]{svjour}
\usepackage{graphics}
\usepackage{graphicx}
\usepackage{dcolumn}
\usepackage{bm}
\usepackage{subfigure}
\usepackage{dsfont}
\usepackage{comment}
\usepackage{amsmath}
\usepackage[colorlinks=true, linkcolor=blue, citecolor=blue, urlcolor=blue]{hyperref}

\usepackage[backend=biber,sorting=none,style=numeric-comp]{biblatex}
\DeclareFieldFormat{pages}{#1}
\DeclareFieldFormat{pagetotal}{#1}
\renewbibmacro{in:}{%
  \ifentrytype{article}
    {}
    {\bibstring{in}%
     \printunit{\intitlepunct}}}
\bibliography{apssamp}
\begin{document}
\title{Thermal effects on stellar neutron capture reactions: a quantum dynamical approach}
\author{Nicholas Lightfoot\inst{1}, Alexis Diaz-Torres\inst{1} \and Paul Stevenson\inst{1}
}                     
\institute{Department of Physics, University of Surrey, Guildford GU2 7XH, Surrey, United Kingdom}
\date{Received: date / Revised version: date}
%
\abstract{The neutron capture process plays a vital role in creating the heavy elements in the universe. Astrophysical environments involved in these processes are characterized by two distinct reaction mechanisms: the slow and rapid neutron capture processes. In this work, the slow neutron capture process is described with the time-dependent coupled channels wave-packet (TDCCWP) method that uses both a many-body nuclear potential and an initial temperature-dependent state to account for the thermal environment.  To evaluate the role of a mixed and entangled initial state in the temperature-dependent neutron capture cross section, TDCCWP calculations are compared with those from the coupled-channels density matrix (CCDM) method based on the Lindblad master equation. The importance of including temperature in the initial wave-function of the TDCCWP approach is compared to a thermalisation of the reaction rate using a Hauser-Feshbach style approach. TDCCWP calculations indicate a decrease of the n+$^{188}$Os capture cross section with increasing temperature, along with a decrease in reaction rates for the highest thermal energies studied, which are contrary to Hauser-Feshbach calculations and important in the rapid neutron capture process. The physical reason for this discrepancy is the key role of the dynamical nuclear coupling between the thermally populated states of the target nucleus, which is neglected in the Hauser-Feshbach approach, but creates a dominant neutron capture pathway with increased neutron speed and thus reduces the neutron capture cross section.
}
\PACS{
      {PACS-key}{discribing text of that key}   \and
      {PACS-key}{discribing text of that key}
     } 
%
%
\titlerunning{Thermal effects on stellar reaction neutron capture reaction rates}
\authorrunning{N. Lightfoot, A Diaz-Torres, P. Stevenson}
\maketitle
\section{\label{sec:level1}Introduction}

The neutron capture reaction is crucial in the creation of heavy nuclei in the universe.  These reactions are understood through reaction cross sections.  Neutron capture reactions in heavy nuclei are split into several categories, depending on the reaction environment.  For example, the slow neutron capture process (s-process) generally occurs at 50-300 million kelvin (5-30 keV in terms of thermal energy) and is part of the helium burning process in stars \cite{reifarth2010s}.  However, the rapid neutron capture process (r-process) occurs in the range of 1-100 billion Kelvin (0.1-10 MeV in thermal energy) and is thought to occur in neutron star mergers \cite{thielemann2017neutron,thielemann2026r}.  Both processes create very neutron rich nuclei beyond iron.  These neutron captures continue until the $\beta^-$ decay rate is more probable than the neutron capture rate.  Once this happens, these neutron rich nuclei "fall" back towards the valley of stable isotopes creating new elements with larger atomic numbers through $\beta^-$ decays.  The role that temperature plays in the environments of these reactions may be relevant due to both the thermal population of low-lying excited states of the target nuclei and the dynamical coupling between these states \cite{lee2023thermal} during the neutron capture process.    

One specific example of the importance of the neutron capture cross section in osmium is in their use in the Re-Os clock \cite{segawa2007neutron,clayton1964cosmoradiogenic}.  This cosmic chronometer is a decay chain that can predict the age of the universe, based on the abundance ratios between $^{187}$Re and $^{187}$Os.  $^{187}$Re has a $\beta$ decay half-life of 42.3 Gyr.  This half-life is much longer than the current estimated age of the universe, $13.787 \pm 0.020$ Gyr \cite{planck2018ageuniverse}.  Looking at these abundances, the $\beta^-$ decay of $^{187}$Re into $^{187}$Os in the crust of the Earth can give a timescale to the age of the universe.  However, $^{187}$Os is also created by the neutron capture process on $^{186}$Os.  This reaction is dominated by the s-process, which restricts the temperatures at which these reactions generally occur.  Therefore, understanding how to calculate the effect of temperature on neutron capture cross sections in osmium isotopes is vital to giving an accurate understanding of the estimated amount of $^{187}$Os that comes from the $\beta^-$ decay of $^{187}$Re. 

Previously, Hauser-Feshbach calculations have been used to determine cross sections and reaction rates used in the network calculations for the age of the universe estimates.  Thermal effects are included through the use of the Maxwell-Boltzmann speed distribution associated with single-channel neutron capture processes along with Boltzmann factors to account for the thermal population of each internal energy level of the target nucleus at a certain temperature \cite{fujii2010neutron}. However, this description neglects the dynamical coupling of the thermally populated energy levels of the target. Previously, this Hauser-Feshbach methodology has been implemented in several packages such as TALYS \cite{koning2005talys} and NON-SMOKER \cite{rauscher2000astrophysical} and has been incorporated with neutron capture cross sections in the laboratory \cite{mosconi2010neutronI,fujii2010neutron,winters1987maxwellian} to generate stellar enhancement factors and stellar reaction rates \cite{rauscher2012sensitivity}.  In the present work, a Hauser-Feshbach style calculation is included and compared with TDCCWP to highlight the effects of dynamical coupling between thermally excited energy levels. The novel TDCCWP method allows for the inclusion of thermal effects in the initial wave function of the neutron capture reaction. The differences of temperature-dependent reaction rates in a TDCCWP calculation against a Hauser-Feshbach style calculation are discussed in the present work for the n+$^{188}$Os reaction as a test case.    

Couplings between excitations of a target nucleus have been included in static coupled channels calculations, such as in the CCFULL and FRESCO codes \cite{hagino1999program,bustreo2013fresco}, where coupled-channels Schr\"odinger's equations are solved directly for the final state of the system, and the transmission coefficients are extracted from the solution. However, a dynamical method can also be used to understand the changes in the population of the energy levels of the target nucleus throughout the reaction, while including thermal effects in the initial state of the system.  For the first time, a wave-packet is used here instead to model the interactions of the neutron-osmium system. This quantum dynamical approach has previously been applied in studies of low-energy heavy-ion fusion \cite{boselli2015quantifying,diaz2018characterizing,vockerodt2019describing,vockerodt2021calculating,vockerodt2021quantum,lee2022coherence,diaz2024cluster,lee2024friction,thomson2024deuterium,thomson2025laser,CLOSE2025139881}. The incorporation of temperature into the initial state of the n+$^{188}$Os collision plays a role in the dynamical populations of the $^{188}$Os internal states.  Then, the interaction between the $^{188}$Os target and a neutron is described by the use of a wave-packet using the TDCCWP framework.

To begin, this approach will be explained in Section II by exploring how the TDCCWP method is used to create a quantum dynamical reaction model. This will be followed by an explanation of thermal effects and reaction rates, as well as a comparison to existing models, in Section III.  Finally, in Section IV, conclusions will be drawn from these results, and future work will be discussed.

\section{Theory}
\subsection{The TDCCWP method}
The present dynamical model uses a wave-packet to describe the interaction of the neutron-osmium system.  This is done by time-evolving the wave-packet along a radial grid so the dynamics of the system and the populations of different energy levels of the $^{188}$Os target can be understood, before, during, and after the interaction with the nuclear and absorptive potentials.  This method is split into several steps as follows.
\begin{enumerate}
\item Initialize the wave-packet at $t=0$ on the radial grid.
\item Propagate the wave-packet along the grid using the time-evolution operator. Time steps can be observed to understand the fluctuations in populations of the internal energy levels of the $^{188}$Os target.
\item Allow the reflected wave-packet to move back to its original position and extract the transmission coefficients from the model to generate capture cross sections and reaction rates.
\end{enumerate}

The above process is carried out using a modified Chebyshev propagation scheme \cite{tal1984accurate,mandelshtam1995bound,diaz2018characterizing} to apply the exponential time-evolution operator to the wave-packet.  Throughout the propagation, the wave-packet interacts with the nuclear potential, resulting in a partial reflection of the wave-packet.  A phenomenological absorption potential is used to model the target nucleus capturing a neutron.  The nuclear potential is parameterised by a Woods-Saxon (WS) potential fitted to a Hartree-Fock (HF) potential model and is explored in the next section.
\subsection{Hartree-Fock  potential}
Using the Sky3D code \cite{maruhn2014tdhf}, a HF nuclear structure code, a single potential is generated across coordinate space in the x,y, and z coordinates, which calculates the average potential felt by a nucleon due to the action of the rest of the nucleons.  The $^{189}$Os nucleus is used as an input, since this is the compound nucleus of the $^{188}$Os+n neutron capture reaction.  The coupled channels model will be introduced in Section 2.6 to include the internal states of the $^{188}$Os nucleus.  

\begin{figure}
    \centering
    \includegraphics[width=\linewidth,keepaspectratio]{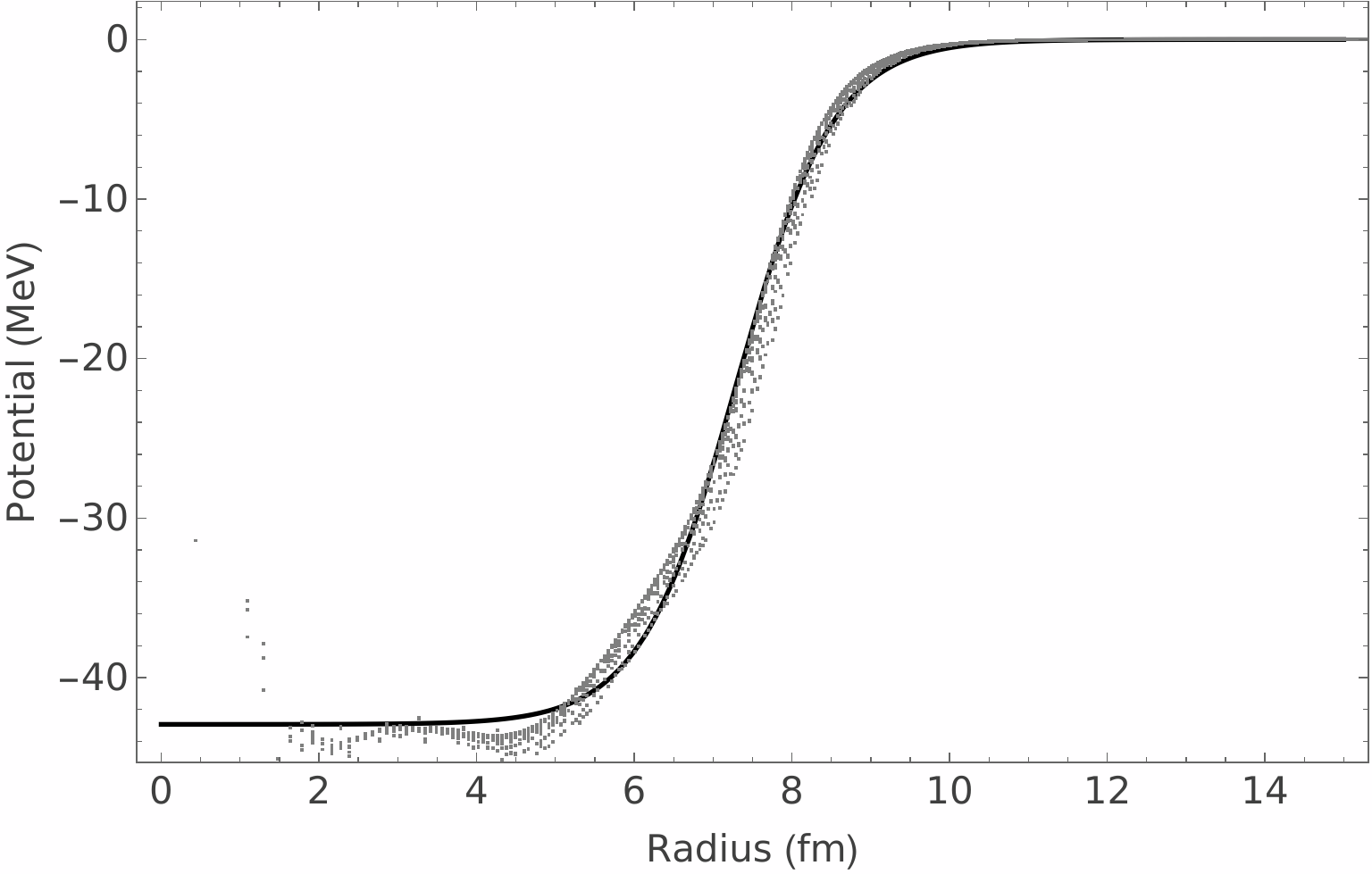}
    \caption{The nuclear interaction potential between a neutron and the $^{188}$Os target is generated from the Sky3D code with a Woods-Saxon shaped line fitted to the data for $^{188}$Os+n using a Newtonian least-squares regression.  The fit is shown in black, with the Hartree-Fock data in gray. The parameterisation data for $^{188}$Os+n is shown in Table \ref{tab:my_label}.}
    \label{fig:RadialWoods Saxon}
\end{figure}

The potentials in Cartesian coordinates are then mapped onto a radial coordinate space, as shown in Fig. \ref{fig:RadialWoods Saxon}.  This potential can be fitted for the parameters, $V_0$, $a_0$, and $R_0$,  using a Newtonian least squares regression, to create an effective Woods-Saxon potential defined as:
\begin{equation}
    V_N=\frac{V_0}{1+\exp(\frac{r-R_0}{a_0})}.
    \label{nuclearpotential}
\end{equation} 
Here $V_0$ is the strength of the potential, $a_0$ defines the diffuseness of the potential and $R_0$ is the radius of the nucleus defined as $R_0=r_0(A_T^{\frac{1}{3}}+A_P^{\frac{1}{3}})$ where $A_{T,P}$ are the atomic numbers for the target and projectile, where for the neutron $A_P$ is 1, and $r_0$ is a radius parameter.

The WS potential is unable to fit the HF potential at small radii because of shell effects in the HF potential.  These shell effects are irrelevant in TDCCWP because of the phenomenological absorption potential and the reflective wall at the origin, converting any negative radii into an equivalent positive radius, an equivalent reflection about the origin.  This boundary is created by the discrete variable representation of the kinetic energy \cite{seideman1992quantum}.  Therefore, the points within 2 fm are omitted, but are shown in Fig. \ref{fig:RadialWoods Saxon} to show the full many-body potential.  The fitted parameters used are shown in Table \ref{tab:my_label}.  As well as the parameters from the Hartree-Fock model the global optical model parameters are shown, which are the parameters used in many reaction models, such as TALYS, which use an energy dependent potential strength.  The maximum and minimum values of the incident energy are selected and shown in Table \ref{tab:my_label}, which are calculated using the Koning-Delaroche parameterisation \cite{koning2003local} or the HF model. The HF potential model is used to implement a many-body microscopic potential.

\begin{table}[hbt!]
    \centering
      \caption{The fitted Woods-Saxon parameters for the static HF potential of $^{188}$Os+n shown in Fig. \ref{fig:RadialWoods Saxon} along with the set of global optical model parameters. The potential strength is energy dependent, so the maximum and minimum incident energy of the neutron are selected.  With the Hartree-Fock (HF) parameterisation and the Koning-Delaroche (KD) parameterisation.}
      
    \begin{tabular}{|c|c|c|c|}
        \hline
         Type & $V_0$ (MeV)& $a_0$ (fm)& $R_0$ (fm) \\ 
        \hline
         HF & $-42.9311$ & $0.6129$ & $7.3033$ \\
         \hline
         KD $E_0=10$ keV& $-48.5707$ & $0.65$ & $7.07$ \\
         \hline
         KD $E_0=500$ keV& $-48.4031$ & $0.65$ & $7.07$ \\
         \hline
    \end{tabular}
    \label{tab:my_label}
\end{table}



\subsection{Initialisation of the wave-packet and system}
To initialise the system, the general form of a Gaussian wave-packet is used to model the initial ($t=0$) radial wave-function of the neutron-osmium system as shown:
\begin{equation}
    \psi(r)=\mathcal{N}^{-\frac{1}{2}} \exp\bigg(\frac{(r-x_0)^2}{2\sigma_0^2}\bigg) \exp(-iK_0r).
    \label{wavepacket0}
\end{equation}    
This uses $\mathcal{N}=\langle\psi|\psi\rangle$ as the normalisation of the wave-function, $x_0$ as the initial position of the wave-packet, and $K_0$ as the wave-number determined by the initial incident mean energy of the wave-packet, $E_0$ \cite{thomson2024deuterium}.  The spatial width of the wave-packet is defined as $\sigma_0$, which also defines the energy resolution of the model.  In other words, a wider wave-packet will give a narrower momentum-space wave-packet from the inverse relation between spatial width and momentum width.  Furthermore, a narrow energy resolution corresponds to a small momentum width and, therefore, a large spatial width, $\sigma_0$.  The fluctuation energy is determined by $\Delta E=\frac{\hbar^2}{4\mu\sigma_0^2}$ \cite{thomson2024deuterium}.  Choosing a larger width requires a larger grid size and, therefore, more grid points for accurate energy resolution.  This then increases the computational time of the model. The selected width of 50 fm is chosen to reduce the computational time with the required resolution, and gives an error in energy of 4 keV.  It can be shown that with a larger width of 80 fm, there is a percentage difference, from the $50$ fm results, of $<1\%$ in transmission coefficients.

\begin{table}[hbt!]
    \centering
    \caption{The initial parameters for the grid size, the wave-packet, and absorptive potential parameters are shown with a description of each parameter used.}
    
    \begin{tabular}{|c|c|c|c|}
        \hline
         Parameter & Value & Description \\ 
         \hline
         N & $400$ & Number of grid points \\
         \hline
         $r_{min}$ & $0.5$ fm & Minimum radius of the grid \\
         \hline
         $r_{max}$  & $500.5$ fm & Maximum radius of the grid \\
         \hline
         $\Delta t$ & $0.1$ zs  & Time step in propagation \\
         \hline
         $x_0$ & $150$ fm & Initial position of wave-packet \\
         \hline
         $\sigma_0$ & $50$ fm & Initial width of wave-packet \\
         \hline
         $\beta_2$ & 0.179 &  $^{188}$Os quadrupole deformation \cite{wang2015nuclear}\\
         \hline
         $\beta_4$ & -0.059 & $^{188}$Os hexadecapole deformation \cite{wang2015nuclear}\\
         \hline
         $V_{W0}$ & -7 MeV & Absorption potential strength\\
         \hline
         $a_{W0}$ & 0.75 fm & Absorption diffuseness\\
         \hline
         $R_{W0}$ & 5 fm & Radius where absorption is 
         centered\\
         \hline 
    \end{tabular}
    \label{tab:parameters}
\end{table}

\subsection{Thermal wave-packet}
The wave-packet can be split into $n$ parts, for each of the internal energy levels of the target.  Each of these components of the wave-packet corresponds to the interaction of the neutron with the nucleus in the given excited state.  The first of these components is the interaction with the ground state, the second is the interaction with the first excited state, and so on. The target's states are in statistical thermal equilibrium with the ambient. While the osmium-neutron radial motion is considered to be in a coherent quantum superposition, the $^{188}$Os target is in a mixed ensemble of states \cite{greiner1997thermo} whose probabilities are given by the Boltzmann distribution. The use of Boltzmann factors is expected to be adequate for nuclear processes at high temperatures, in which excited states of a target nucleus are in thermal equilibrium with the ground state, as explained in Ref. \cite{fowler1967thermalequilibrium}. The general form of the components of the initial thermal wave-packet is:

\begin{equation}
    |\psi\rangle=\sqrt{p_0}|\psi_0\rangle+\sum_{n=1}^N \sqrt{p_n}|\psi_n\rangle.
    \label{thermal_expansion}
\end{equation}
The value of $p_0$ is calculated using the normalisation condition of the wave-packet, $p_0=1-\sum_{n=1}^Np_n$. The thermal population of excited states reads as follows \cite{greiner1997thermo}:
\begin{equation}
    p_n=\frac{(2I_n +1)\exp({-\frac{\epsilon_n}{kT}})}{\sum_{i=0}^I (2I_i +1) \exp({-\frac{\epsilon_i}{kT}})},
    \label{partition}
\end{equation}
where $\epsilon_n$ is the energy of the $n^{th}$ excited state relative to the ground state ($\epsilon_0=0$ MeV), $I_n$ denotes the spin of the state, $k$ is the Boltzmann constant and $T$ is the temperature of the environment. The denominator in Eq. (\ref{partition}) defines the partition function which forces the system into a state where at thermal equilibrium a population inversion is impossible and the ground state can have at least the same probability as the excited states. The spin degeneracy factors in Eq. (\ref{partition}) will be neglected in the present calculations because only excited states of the ground-state rotational band of $^{188}$Os are included. This is because the Coriolis interaction is neglected and only sub-states with the same magnetic quantum number as the ground state are coupled with each other. In Eq. (\ref{thermal_expansion}), the relative phase between the components of the wave-packet is unimportant, as the initial interference in the region of the nuclear coupling at small radii does not contribute significantly to the neutron capture probability. This is demonstrated below by comparing the capture probabilities using the present TDCCWP method and the CCDM approach, which are quantitatively very similar. In the CCDM approach \cite{lee2022coherence,lee2023thermal}, which is based on the Lindblad master equation for a coupled-channel density matrix with an initial density matrix $\rho_0 = \sum_{n=0}^N p_n|\psi_n\rangle \langle\psi_n|$, there is no initial interference between the components of the wave-packet. The present method is an approximate model and provides a more efficient quantum dynamical treatment of thermal effects on neutron capture reactions than the CCDM approach. Applying Eqs. (\ref{thermal_expansion}) and (\ref{partition}) to the case where there are only two channels gives \cite{lee2023thermal}:
\begin{equation}
    |\psi\rangle=\sqrt{1-p_1}|\psi_0\rangle+\sqrt{p_1}|\psi_1\rangle.
    \label{two_state}
\end{equation}


\subsection{Components of the thermal wave-packet and kinematically closed channels}
The initial wave-packet related to each excited state of the $^{188}$Os target, which are the components of the thermal wave-packet in Eqs. (\ref{thermal_expansion}) and (\ref{two_state}), is defined by an expression like Eq. (\ref{wavepacket0}) with the normalisation $\mathcal{N}=p_n$ and an asymptotic mean wave-number, $K_n$, which provides an incident mean energy of $E_0 - \epsilon_n$, $\epsilon_n$ being the excited state energy relative to the ground state.   Importantly, the wave-packets described in the previous section are the interactions between the neutron and the internal energy levels allowing for the difference in kinetic energies in the initial state.  For instance, this mean wave-number for an s-wave neutron is given by:
\begin{equation}
    K_n=\sqrt{\frac{2\mu(E_0-\epsilon_n)}{\hbar^2}-\frac{1}{2\sigma_0^2}},
\end{equation}
where the fluctuation energy of the wave-packet, $\frac{1}{2\sigma_0^2}$, is included \cite{thomson2024deuterium}.  When $E_0 < \epsilon_n$, computing $K_n$ requires a negative square root due to a negative value for the incident mean energy. Consequently, the first excited state of the $^{188}$Os target can only be virtually excited by nuclear coupling in the course of the collision and thus contributes to the probability of neutron capture. Any probability in the first excited state is "stuck" inside the nucleus and does not move along the radial grid at all. This scenario corresponds to kinematically closed channels in the $^{188}$Os-n system, in which we set $K_n=0$ fm$^{-1}$.

\subsection{Hamiltonian for the coupled channels method}
In order to include the energy levels of $^{188}$Os in the Hamiltonian, the coupled channel framework is implemented. The Hamiltonian must now be extended to account for the interactions of the neutron with the target relating to the ground and first excited states. The ground state Hamiltonian is described as:

\begin{equation}
    \hat{H}_0=-\frac{\hbar^2}{2\mu}\frac{d^2}{d r^2}+\frac{\hbar^2l(l+1)}{2\mu r^2}+V_N(r)+iW(r).
\end{equation}
The kinetic energy operator is treated using the Discrete Variable Representation (DVR) method \cite{seideman1992quantum}, which prevents unphysical 'wrap-around' effects at the edges of the radial grid, and the second term is the centrifugal potential. The $V_N(r)$ defines the nuclear interaction potential of the system, and the $W(r)$ is the absorption potential.  The complete coupled channel Hamiltonian can be described by a matrix as shown below:
\begin{equation}
    \hat{H} |\psi\rangle=\begin{pmatrix} \hat{H}_0+V_{0-0} & V_{0-2}\\V_{2-0}&\hat{H}_0+V_{2-2}+\epsilon_2\end{pmatrix}\begin{pmatrix}|\psi_0\rangle\\|\psi_2\rangle\end{pmatrix}.
\end{equation}
Here and for many other even-even nuclei, $0^+$ is the ground state and $2^+$ is the first excited state, and so referred to as 0 and 2 throughout.  These labels specify the ground state and the first excited state, respectively, for the $^{188}$Os ground-state rotational band. For odd mass nuclei and odd-odd nuclei, this process is slightly different \cite{thompson2014coupled}. The first entry in the Hamiltonian, 0-0, is identical to the case without coupling and corresponds to the ground state, while the final entry, 2-2, is the Hamiltonian of the first excited state, which contains the interaction of the neutron with the first excited state along with the energy shift of this state, $\epsilon_2$.  Furthermore, the off-diagonal entries, $V_{0-2}$ and $V_{2-0}$, give the components of the Hamiltonian relating to couplings between the ground state and the first excited state.  The wave-packet also requires components corresponding to the ground state and first excited state interactions.  If there were more than 2 states, this would then be generalised with a $n\times n$ matrix with the number of excited states, $n-1$, as well as a wave-function with $n$ parts \cite{vockerodt2019describing}.  However, 2 states are used here since there is an insignificant difference, $<$ 0.1\%, in the cross sections between 2 and 3 states, and the use of 2 states greatly reduces computational time.  This could be another area that defines the error of the model; however, it is generally a smaller contribution than increasing the width parameter to reduce the energy width.  The calculation of these coupling potentials is now shown in the next section.
\begin{figure}[h]
    \centering
    \includegraphics[width=\linewidth]{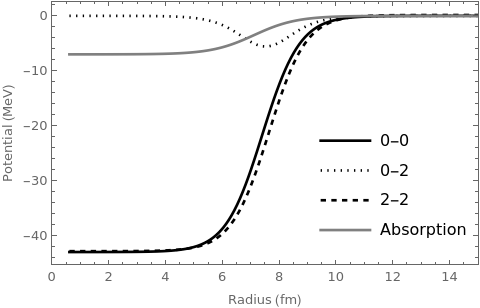}
    \caption{All of the different potentials in the model for the $^{188}$Os+n reaction are presented, including the coupling potential.  Note the positioning of the absorption potential, which is more central than the coupling potential and well inside the nuclear interaction region.  The 0-2 shows the coupling between the ground and first excited state.}
    \label{fig:couplechanpot}
\end{figure}

\subsection{The Coupled Channels Potential}
The above effect of couplings on the Hamiltonian is implemented through a displacement of the nuclear potential for each of the excitations.  Moreover, this coupled-channel approach creates a new potential that describes the coupling among energy levels of the interacting system. This coupling potential is shown below, where $\hat{r}_{cc}$ is a dynamical operator of the coupling matrix that defines the displacement of the potential due to the couplings as follows: 
\begin{equation}
    V_{cc}=\frac{V_0}{1+\exp((r-R_{0}-\hat{r}_{cc})/a_{0})}.
\end{equation}
 This new form of the potential can then be used in order to define the effects of excitations and de-excitations from one channel to the next.  The general form of the total coupling potential is \cite{hagino1999program}:
\begin{align}
    \label{coupling equation}
    V_{N,nn'}&=\langle I_n|V_N(r,\hat{r}_{cc})|I_{n'}\rangle-V_N(r,0)\delta_{nn'}\\
    &=\sum_\alpha\langle I_n|\alpha\rangle\langle\alpha|I_0'\rangle V_N(r,r_{cc,\alpha})-V_N(r,0)\delta_{nn'}.\nonumber
\end{align}

In the above $\hat{r}_{cc}|\alpha\rangle=r_{cc,\alpha}|\alpha\rangle$ where the eigenstates $|\alpha\rangle$ are created from the basis spin states $|I_n\rangle$, which define the energy levels of $^{188}$Os.  The second term, which includes the delta function, is used for the case of the potentials for the ground and first excited states coupling to themselves and avoids double counting of the original nuclear potential $V_N$ from above. These eigenstates are found by diagonalising the following coupling matrix \cite{hagino1999program}:
\begin{equation}
    \langle I_n|\hat{r}_{cc}|I_{n'}\rangle=R_{cc}(\beta_2 F(2,I_n,I_n')+\beta_4 F(4,I_n,I_n')).
    \label{coupling matrix}
\end{equation}
Here, $\beta_2$ and $\beta_4$ are the quadrupole and hexadecapole deformation parameters that define the shape of the target nucleus.  These are shown in Table \ref{deftable}.  The $R_{cc}$ is the phenomenological value of the coupling radius defined as $R_{cc}=r_{coup} (A_T)^\frac{1}{3}$ fm where $r_{coup}$ is the coupling radius parameter.  The function $F$ here is defined in the following as the form factor, where the bracketed portion is a 3-j symbol.

\begin{align}
    F(I,I_n,I_{n'})=&\sqrt{\frac{(2I+1)(2I_n+1)(2I_{n'}+1)}{4\pi}}\\&\times\begin{pmatrix}I_n & I & I_{n'}\\0 & 0 & 0 \end{pmatrix}^2.\nonumber
\end{align}
For a system of $n$ energy levels, there will be $n$ different coupling eigenvalues from Eq. (\ref{coupling matrix}) that are used with Eq. (\ref{coupling equation}) to calculate the full coupling potentials.  The $0^+$ and $2^+$ states are used here.  This coupling channel scheme has commonly been used in static coupled-channel frameworks, such as those implemented in the CCFULL \cite{hagino1999program} and FRESCO \cite{thompson1988coupled} codes.  In the next section, a description of the dynamical extension of this coupled-channel model is shown, along with the implementation of the absorption potential.



\begin{table}[hbt!]
    \caption{Deformation from different theoretical calculations namely the Potential Energy Surface (PES) model \cite{wang2015nuclear}, Folded-Yukawa Finite-Range Droplet Model (FY+FRDM) \cite{moller2016nuclear}, and the Extended Thomas-Fermi plus Strutinsky Integral (ETFSI) method \cite{aboussir1995nuclear}.  The experimental estimate calculated from reduced transition probabilities is $\beta_2=0.186$ \cite{wang2015nuclear}}
    \centering
    \begin{tabular}{|c|c|c|}
        \hline
         Model & $\beta_2$ & $\beta_4$ \\ 
        \hline
         \textbf{PES}& \textbf{0.179} & \textbf{-0.059} \\
         \hline
         FY $+$ FRDM& $0.192$ & $-0.086$ \\
         \hline
         ETFSI & $0.20$ & $-0.08$ \\
         \hline
    \end{tabular}
    \label{deftable}
\end{table}


\subsection{The modified Chebyshev method for absorptive dynamics}
 
To understand the probability that a neutron being captured by the nucleus, a phenomenological absorption potential is used.  This Woods-Saxon-shaped potential, as defined by Eq. (1), acts to reduce the normalisation of the incident wave-packet. Some care is needed for the placement of this potential, as channel couplings need to take effect inside the nucleus before the absorption potential reduces the norm of the wave-packet \cite{hodgson1984neutron}.  The positions of these potentials can be seen in Figure \ref{fig:couplechanpot}.  A potential too close to the edge of the nuclear radius would not allow any effects from the couplings, whereas a potential too close to the center of the nucleus will not maximally absorb normalisation.  This location, described by the parameters in Table \ref{tab:parameters},  still gives maximum absorption with $R_{W0}=5$ fm, but still allows the couplings to take place.  Similarly, the absorption strength of $V_{W0}=-7$ MeV is sufficiently large to have a $<0.1\%$ change in cross section if a larger $V_{W0}=-50$ MeV absorption strength was chosen, showing the independence of the cross section on these absorption parameters.

With the inclusion of this absorption potential, the modified Chebyshev propagator is used in the TDCCWP method. The Chebyshev propagator rewrites the exponential time evolution operator as a set of Chebyshev polynomials \cite{chen1999chebyshev}.  In general, using an absorption potential in the Hamiltonian would make the Hamiltonian non-Hermitian. To deal with this, an operator is used to reduce the normalisation of the wave-packet while leaving the Hamiltonian real.  This method applies a recurrence relation in which the absorption potential acts as the wave-packet propagates in time using Chebyshev polynomials \cite{mandelshtam1995bound}.  The complete method can be found in \cite{mandelshtam1995bound} and \cite{diaz2018characterizing}.  This method gives the basis for the time propagation and calculation of both the capture probabilities and associated cross sections.


\subsection{Transmission coefficients and the neutron capture cross section}
The TDCCWP implementation ends once the wave packet has fully interacted.  This is defined when the expectation value of the position of the reflected wave-packet reaches the initial position of the incoming wave-packet.  The transmission coefficients are found directly from the reflection coefficients, $\mathcal{R}$, using the relation $\mathcal{T}=1-\mathcal{R}$.  In the case with multiple energy levels, both components of the wave-function are summed since both components of the wave-packet lose normalisation from the absorption.  Furthermore, the calculation of the cross-section is:
\begin{equation}
    \sigma(E_0,T)=\frac{\pi \hbar^2}{2\mu E_0}\sum_{l=0}(2l+1)\mathcal{T}(E_0,l, T).
    \label{crosssection}
\end{equation}
It can be seen that the transmission coefficients depend on the incident energy, $E_0$, as well as the temperature, $T$, of the environment.  The sum over angular momentum in Eq.(\ref{crosssection}) at low incident energies can typically be omitted due to high centrifugal barriers for non-zero partial waves.  The sum over angular momentum reflects more of the wave-packet since the incident energies in neutron capture reactions are generally small, $<$1 MeV.  The next step is to understand how these capture cross sections can be used to calculate reaction rates.

\subsection{Calculating reaction rates}
The reaction rate is defined as the number of reactions that occur per unit of time per unit volume.  In typical calculations, thermal effects are applied, as the reaction rates are calculated using the Maxwell-Boltzmann speed distribution and Boltzmann factors to give probability weightings for thermal effects at each incident energy.  However, in this TDCCWP model, the thermal dependence in the reaction rates will be included at the initialisation of the wave-packet.  The reaction rate in terms of the new thermal dependence of the cross section is calculated as follows \cite{martin2019nuclear,rolfs1988cauldrons}:
\begin{equation}
    \langle \sigma v \rangle=\sqrt{\frac{8}{\pi \mu}} \bigg(\frac{1}{kT}\bigg)^{\frac{3}{2}} \frac{\int_{E_i}^{E_f}{\sigma(E_0,T)\cdot E_0 \cdot P(E_0,T)} dE_0}{\int_{E_i}^{E_f}{E_0 \cdot P(E_0,T)} dE_0}.
    \label{reactionrates}
\end{equation}

A simple substitution with velocities results in an integral over incident energies; in this calculation, a range of $E_i=10$ keV to $E_f=1500$ keV is used.  This also defines $P(E_0,T)$ as the Maxwell-Boltzmann distribution of neutron incident energies.  This substitution gives rise to the pre-factor shown above \cite{martin2019nuclear}.  It is important to note that the original form of this equation \cite{rolfs1988cauldrons} takes the integral from $[0,\infty)$.  This would be unphysical here, so to restrict the calculation to a finite range of incident energies, a normalisation factor is included as the denominator in Eq. (\ref{reactionrates}). The cross-section is given by Eq. (\ref{crosssection}).  In the next section, the results of this temperature dependence are shown against previously used models and against the same model, one with a temperature dependence and one without.

\section{Results}
\subsection{A comparison of transmission coefficients between TDCCWP and CCDM}

\begin{figure}[t!]
    \centering
    \includegraphics[width=\linewidth]{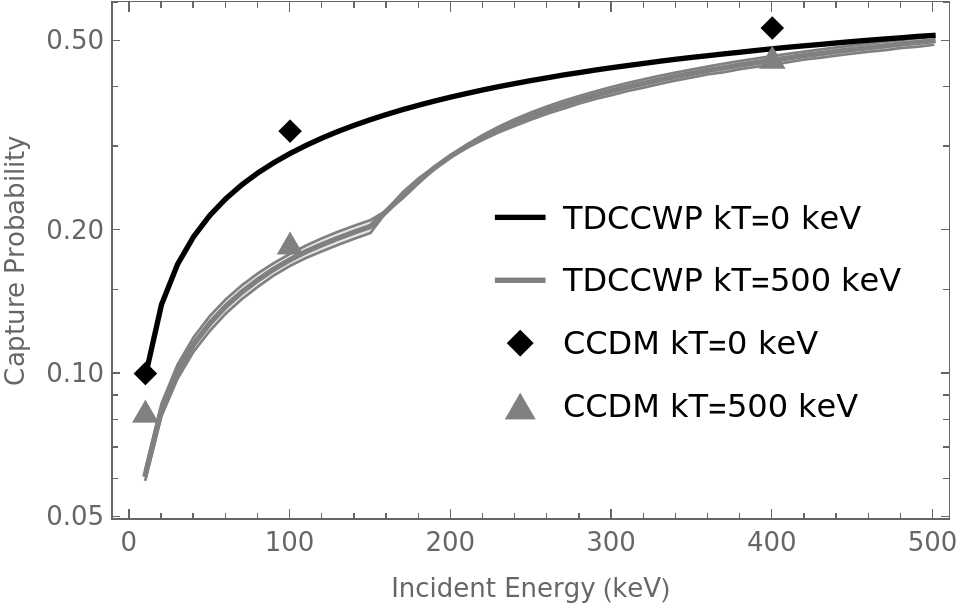}
    \caption{Transmission coefficients from TDCCWP calculations are compared to those from the CCDM model for the $l=0$ partial wave for the $^{188}$Os+n reaction.  Two thermal energies are displayed, these being 0 and 500 keV in thermal energy (kT).  The shaded region around the 500 keV TDCCWP results shows the error in the TDCCWP calculation found by finding the percentage change from a wider wave-packet calculation.}
    \label{fig:ccdmcomp}
\end{figure}

A comparison between TDCCWP and the quantum dynamical CCDM method is useful. CCDM has already been validated to match static coupled-channel calculations \cite{lee2023thermal,vockerodt2021calculating}, so validate the TDCCWP method simultaneously. CCDM also uses a density matrix formalism to explore decoherence effects dynamically, so we here compare a calculation with and without the inclusion of decoherence effects. This has been used to implement thermal effects on heavy-ion fusion reactions to generate thermally dependent cross sections \cite{lee2022coherence,lee2023thermal}. The comparison in Fig. \ref{fig:ccdmcomp} shows the difference between mixed and entangled initial states as well as the thermal dependence of neutron capture cross sections.  

Irrespective of the neutron incident energy, the static coupling between the two channels creates two decoupled pathways for the neutron-osmium interaction \cite{lee2023thermal}. These pathways are the eigenstates of the coupling potential matrix, which have separate statistical weights that depend on temperature. As the temperature increases, the population of the pathway with the most attractive nuclear interaction increases, while the population of the pathway with the least attractive nuclear interaction decreases \cite{lee2023thermal}.  Each of these pathways is a coherent linear superposition of the $^{188}$Os states.  Furthermore, each pathway contains the contribution of the target's ground state, irrespective of the incident energy of the neutron. Therefore, in the case where there is an increase in temperature, the neutron's speed increases in the absorption region.  This increase in speed results in a higher probability for the neutron to escape its capture and therefore a decrease in the capture probability, as observed in Fig. \ref{fig:ccdmcomp}.

The difference in transmission is obvious between both a density-matrix and a wave-packet-based formalism, particularly for small incident energies where coherence effects may play a larger role.  It can be seen that, when generating a transmission coefficient, there is a difference in the calculation below and above the energy of the first excited state (155 keV).  A decrease in the transmission can also be seen in the lower energies, reflecting this change.  At higher energies, the two converge, reflecting what would be expected for kinematically closed and open channels.  In other words, these components can now fully react with the nucleus and are absorbed, whereas the $^{188}$Os target being virtually excited when the incident energies are in the kinematically closed region. Despite this, the two quantum dynamical models provide similar transmission coefficients, particularly for the temperature-independent calculation, $kT=0$ keV. However, the TDCCWP method is computationally more efficient.

\subsection{Sensitivity to the interaction nuclear potential and the deformation parameters}

\begin{figure}
    \centering
    \includegraphics[width=\linewidth]{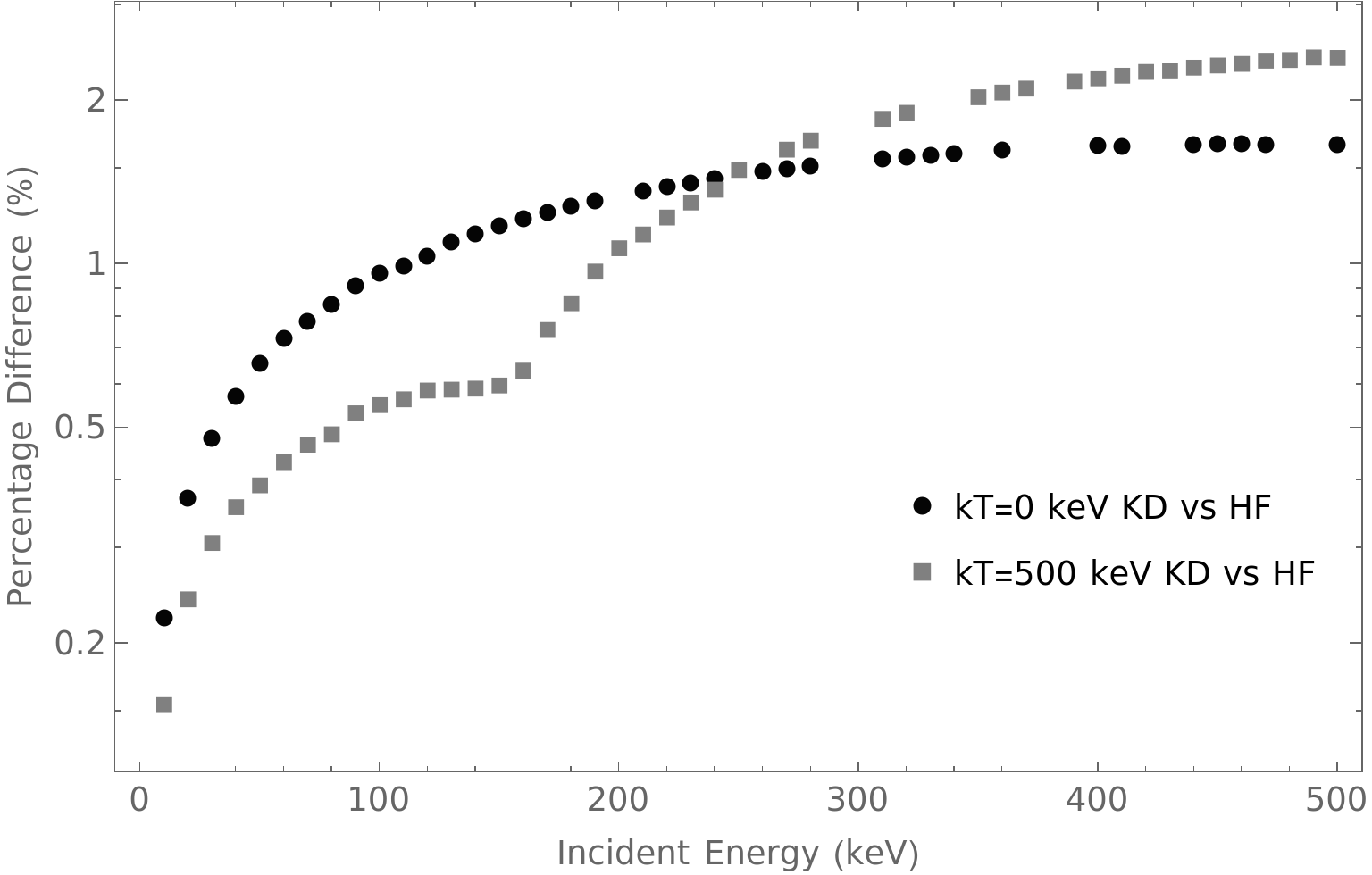}
    \caption{The percentage difference, $100\times\frac{T_{KD}-T_{HF}}{T_{HF}}$, of the $l=0$ transmission coefficients calculated with Koning-Delaroche (KD) and HF nuclear potentials, for thermal energies $kT=0$ keV and $kT=500$ keV. The potential parameters are given in Table \ref{tab:my_label}. A difference of 2\% in transmission coefficients is shown for the largest incident energies.}
    \label{percentage_difference}
\end{figure}

A comparison of the percentage difference between the capture probabilities calculated with the Koning-Delaroche nuclear potential against the HF nuclear potential, whose values are given in Table \ref{tab:my_label}, is shown in Fig. \ref{percentage_difference}. It can be seen that at the highest energies a difference of 2\% in transmission coefficients is seen from the HF values.  As the thermal energy of the model increases, this percentage difference increases, showing an important relationship between the nuclear potential parameters and the capture probabilities. However, the difference between the energies used in the Koning-Delaroche potential results in changes in transmission coefficients of less than 1\%. This means that using $E_0=10-500$ keV does not result in a large difference in transmission coefficients.

\begin{figure}
    \centering
    \includegraphics[width=\linewidth]{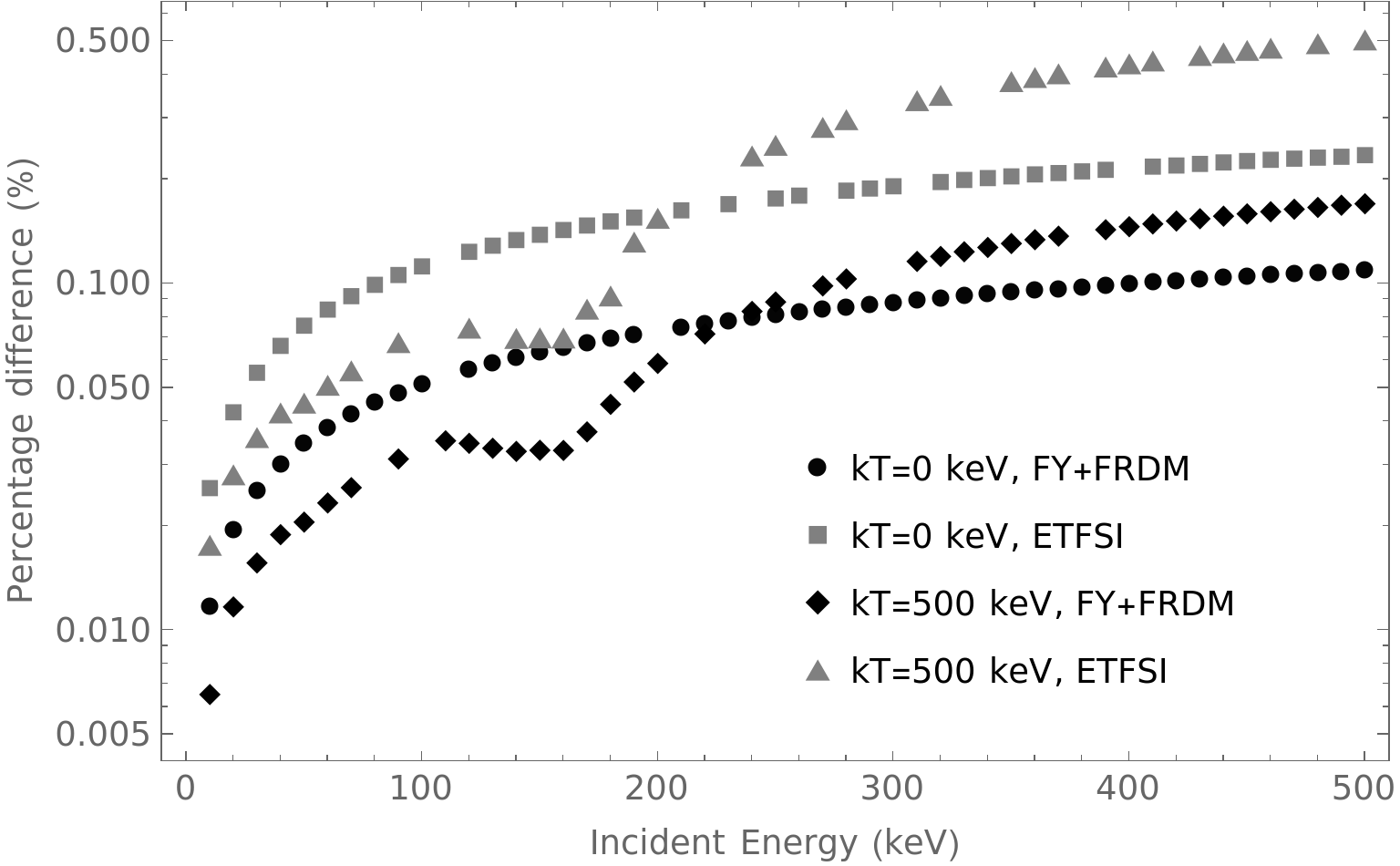}
    \caption{The percentage difference of $l=0$ transmission coefficients, $100\times\frac{T_{FY/ET}-T_{PES}}{T_{PES}}$, of the FY+FRDM and ETFSI deformation models against the PES deformation model, using the HF nuclear potential, at both $kT=0$ keV and $kT=500$ keV. The largest percentage difference is ~0.2\% for the $kT=0$ keV, which is insignificant.  A thermal energy of $kT=500$ keV is also shown with a maximum percentage difference of 0.5$\%$ for the highest incident energies used.  }
    \label{defpercentagedif}
\end{figure}

The deformation parameters describe the intrinsic shape of the nucleus. Several different methods exist to calculate these and are shown in Table \ref{deftable} with the values used in the TDCCWP calculation of the reaction $^{188}$Os + n in bold. The impact of changing the deformation values on the calculated transmission coefficients in TDCCWP can be seen in Fig. \ref{defpercentagedif}, where the differences in the transmission coefficients are less than 1\% for any change in the set of deformation parameters.

\subsection{Cross Sections}
\begin{figure}[t]
    \centering
    \includegraphics[width=\linewidth]{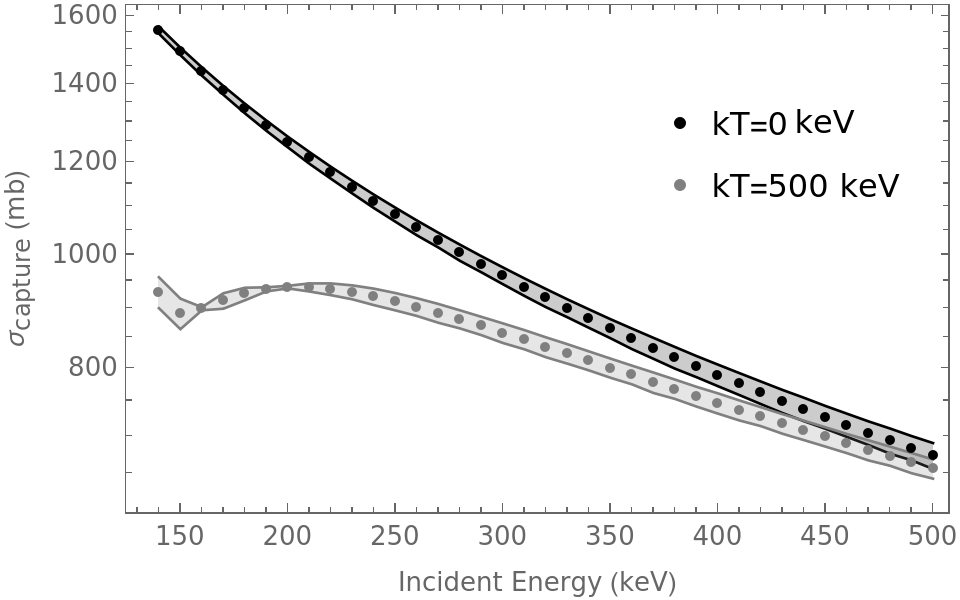}
    \caption{Temperature-dependent capture cross sections for a l=0 neutron on the $^{188}$Os target as a function of the neutron incident energy in the range 140-500 keV. The shaded regions highlight errors associated with the energy variance of the neutron energy (i.e., the differences in the TDCCWP calculations using a spatial width of the initial wave-packet of 50 and 80 fm). The change from the kinematically closed energy region to the open one happens at the $^{188}$Os first excited state energy (155 keV).}
    \label{fig:150-500therm}
\end{figure}

\begin{figure}[h]
    \centering
    \includegraphics[width=\linewidth]{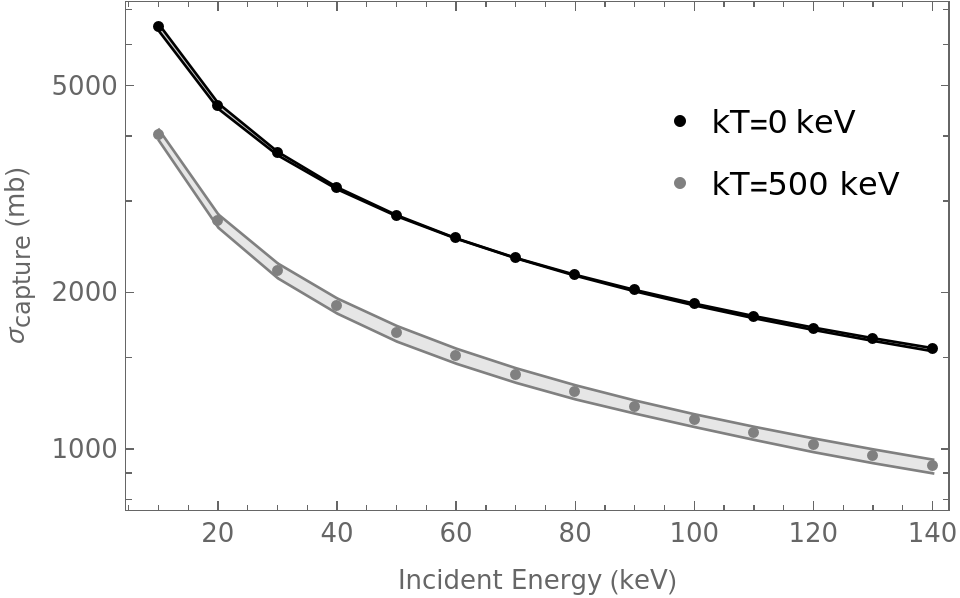}
    \caption{The same as in Fig. \ref{fig:150-500therm}, however, the range of incident energies is 10-140 keV.}
    \label{fig:10-150therm}
\end{figure}

To explore more of this difference across the two incident energy regions, the neutron capture cross sections can also be calculated.  Fig. \ref{fig:150-500therm} shows a kink in the cross-section data where the model transitions from the kinematically closed region, below 155 keV, where there are only virtual excitations to the first excited state, to the region above this where excitations to the first excited state can occur.  For higher incident energies, the higher thermal energy and low thermal energy cases both begin to converge to similar values.  The resonance region for this reaction is below this first excited state and can be seen in Fig. \ref{fig:10-150therm}, again showing the discrepancy in capture cross sections when thermal effects are included.  The thermal energy of 250 keV (5.8 GK)  is chosen to show the large discrepancy when thermal effects are included at the temperature scale of the rapid neutron capture process.  These differences persist but are smaller for the temperatures corresponding to the slow neutron capture process (0.1-1 GK), which is the process that dominates for this isotope \cite{avila2011tungsten}.

\begin{figure}[b]
   \centering
    \includegraphics[width=\linewidth]{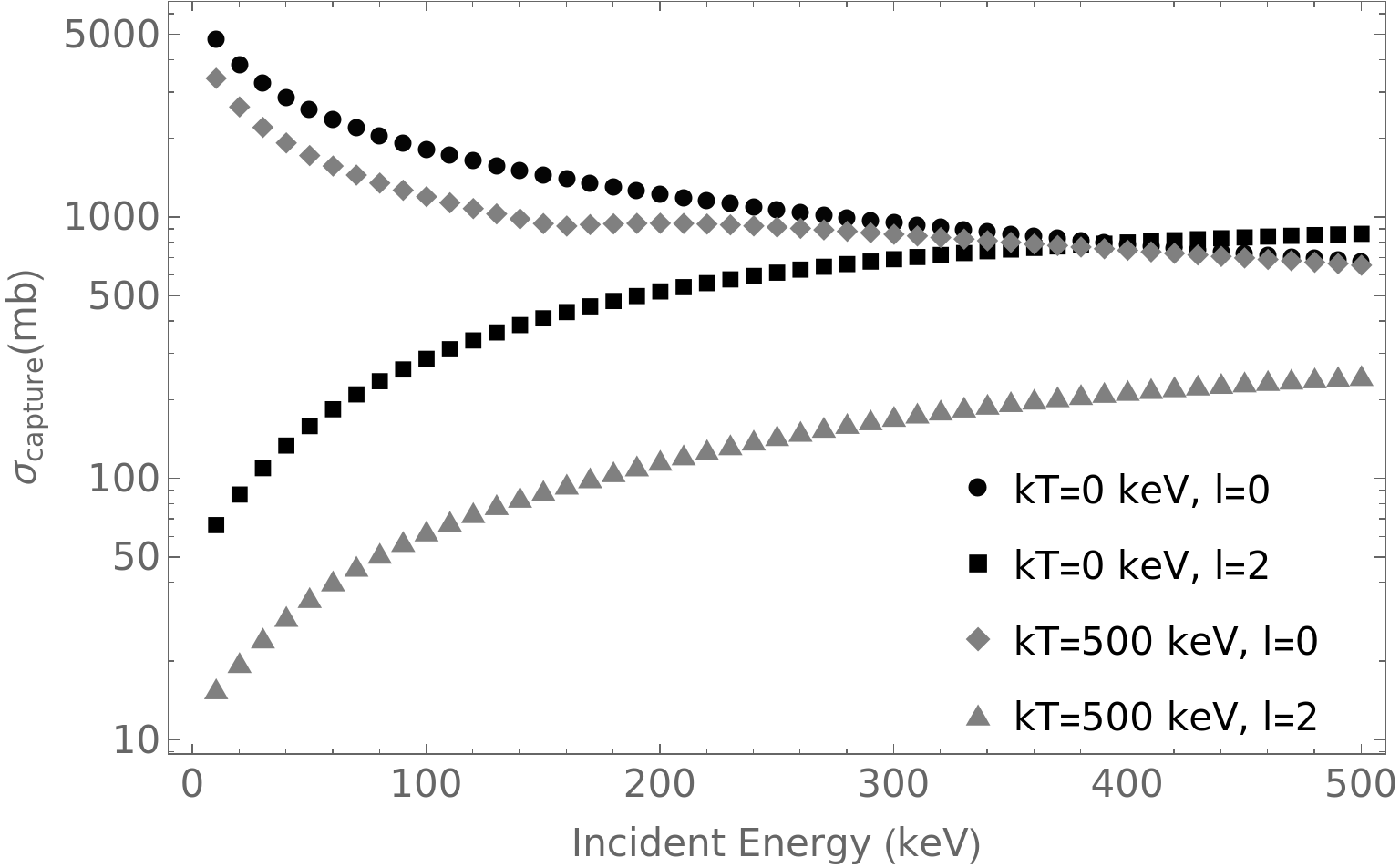}
    \caption{A comparison between the capture cross sections in the $l=0$ and $l=2$ angular momentum states.  With each higher angular momentum, the centrifugal barrier reflects more of the incident wave-packet, resulting in the ability to truncate the higher angular momentum components of the cross-section.  This truncation will occur at the angular momentum, determined by when the component of cross-section is $<1\%$ of the total cross-section.  This cut-off depends on the incident energy being large enough to overcome the height of the centrifugal barrier at a certain angular momentum.}
   \label{fig:higher}
\end{figure}

This comparison can be continued for the higher angular momenta, as shown in Fig. \ref{fig:higher}.  It can be seen that the contributions at the lower incident energies are much less for the higher angular momenta, giving the convergence of the system.  For many of the energies below the first excited state and in the region of importance for resonances, the $l=2$ angular momentum contribution is less than a tenth of the total capture cross section.  This can be continued and shown to converge as the angular momentum increases.  However, for many of these low incident energies, the $l=2$ component is the highest that needs to be included to show convergence of the cross-section.


\subsection{A comparison between theory and experiment}

\begin{figure}[hbt!]
    \centering
    \includegraphics[width=\linewidth]{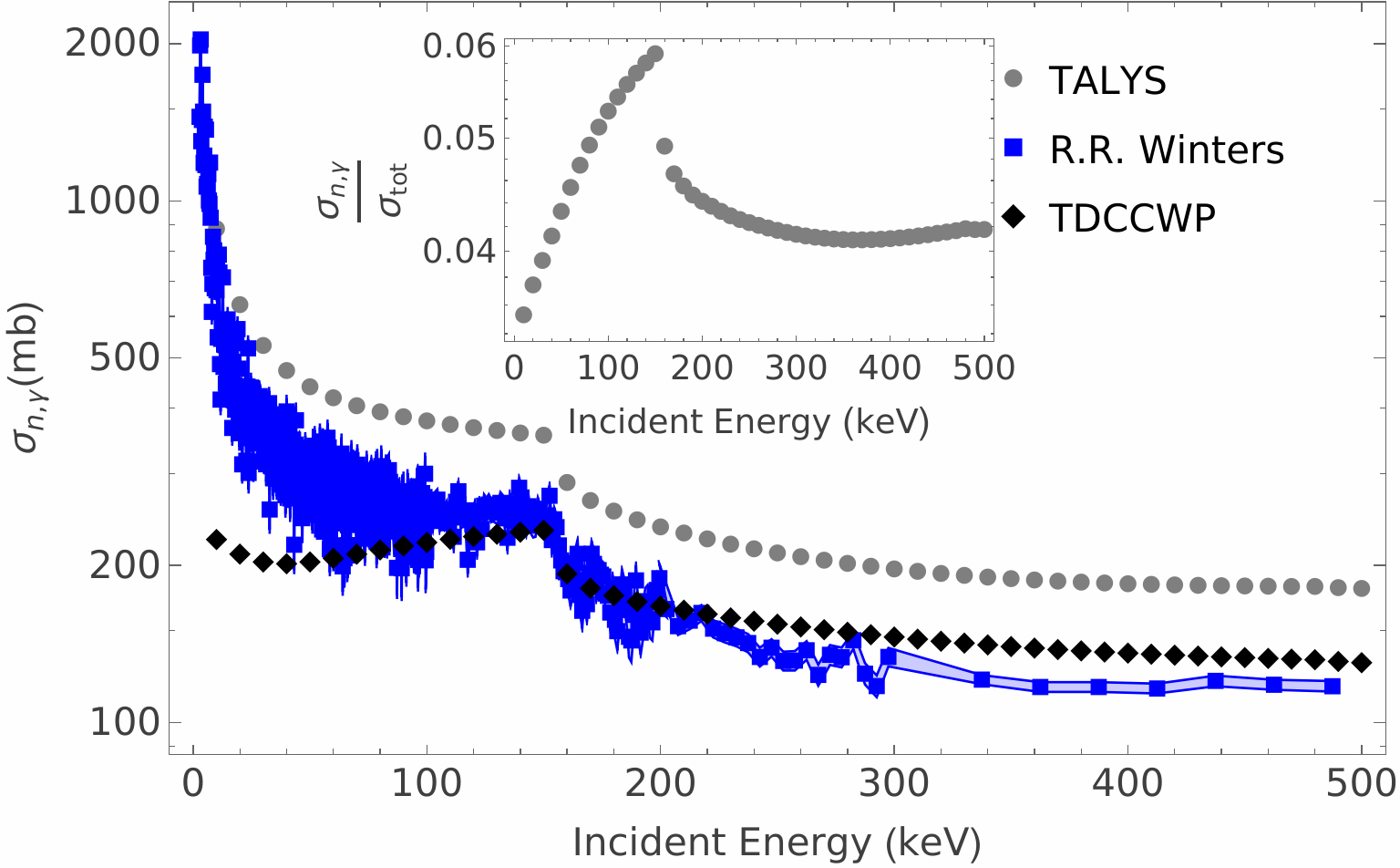}
    \caption{Temperature-independent (n,$\gamma$) cross-sections are shown for the TDCCWP calculation, the experimental data from R. R. Winters \cite{winters1987maxwellian}, and a Hauser-Feshbach style calculation using TALYS \cite{koning2005talys}.  The TDCCWP calculation is originally a total neutron capture cross section calculation with a scaling factor calculated from the ratio between the TALYS calculation of the total neutron capture and (n,$\gamma$) cross sections.  The inset shows the scaling calculated using the (n,$\gamma$) and total neutron capture cross sections.}
    \label{fig:TDCCWPvsEXPvsTALYS}
\end{figure}

Experimental data for the $^{188}$Os+n reaction are prevalent for low incident energies, which makes a comparison of the experimental data and the TDCCWP approach paramount.  However, the available experimental cross sections are the (n,$\gamma$) cross sections, which are a component of the total neutron capture cross-sections, which are calculated in TDCCWP. Therefore, a scale must be applied to the TDCCWP cross-sections to describe the (n,$\gamma$) cross sections. TALYS allows for the calculation of both the total neutron capture cross-sections and the (n,$\gamma$) cross sections. Using the same parameters for the nuclear potential, as in Table \ref{tab:my_label}, a TALYS calculation can give an estimated scaled TDCCWP result of the (n,$\gamma$) cross-sections.  Since, the evaporation process is statistical in nature the fraction of (n,$\gamma$) to total neutron capture cross-section should be reliable in the Hauser-Feshbach model used in TALYS.  The fraction of the total neutron capture cross section to the (n,$\gamma$) cross section in TALYS is then calculated and applied to each incident energy in TDCCWP.  This produces Fig. \ref{fig:TDCCWPvsEXPvsTALYS}, where a fairly good agreement in cross-sections can be seen for the incident energies above 50 keV. It is important to note that the TDCCWP calculations are not fitted to any experimental data. The TALYS calculation, which is a Hauser-Feshbach style calculation, is also not fitted to the experimental data.  

A true comparison of an equivalent (n,$\gamma$) cross section for TDCCWP is found here, which requires an identical parameter set to the Hartree-Fock nuclear potentials to be used in the scaling with TALYS as was used in the TDCCWP calculation.  This results in a large disagreement with the TALYS results to the experimental data in Fig. \ref{fig:TDCCWPvsEXPvsTALYS}, because the calculation is not fitted directly to fit the data.  This data can be found using the TENDL nuclear data evaluation which fits TALYS calculations to experimental data.  The large difference between the two calculations is the first hint that the two calculations, TDCCWP and a Hauser-Feshbach, are not equivalent calculations.  However, the Hauser-Feshbach should be reliable in calculating these scaling factors between channels because of the statistical decay of the compound nucleus.  For the rest of this work the total neutron capture cross section alone is used to compare the two implementations of the neutron capture reactions. 

The next step is to explore how a calculation of reaction rates is changed when cross sections with temperature included at the initialisation of the wave-packet are used, instead of the general inclusion of thermal effects after cross sections have been calculated.

\subsection{Reaction Rates}

The reaction rate is specifically applicable to astrophysical data analysis, where accurate reaction rates give a sense of how experimental cross sections can be mapped onto astrophysical data.  In the neutron capture reaction, this means that accurate reaction rates describe how likely these reactions are to take place at different temperatures, ultimately highlighting the importance of the temperature of the environment on the neutron capture reaction as a whole.  

\begin{figure}[h]
    \centering
    \includegraphics[width=\linewidth]{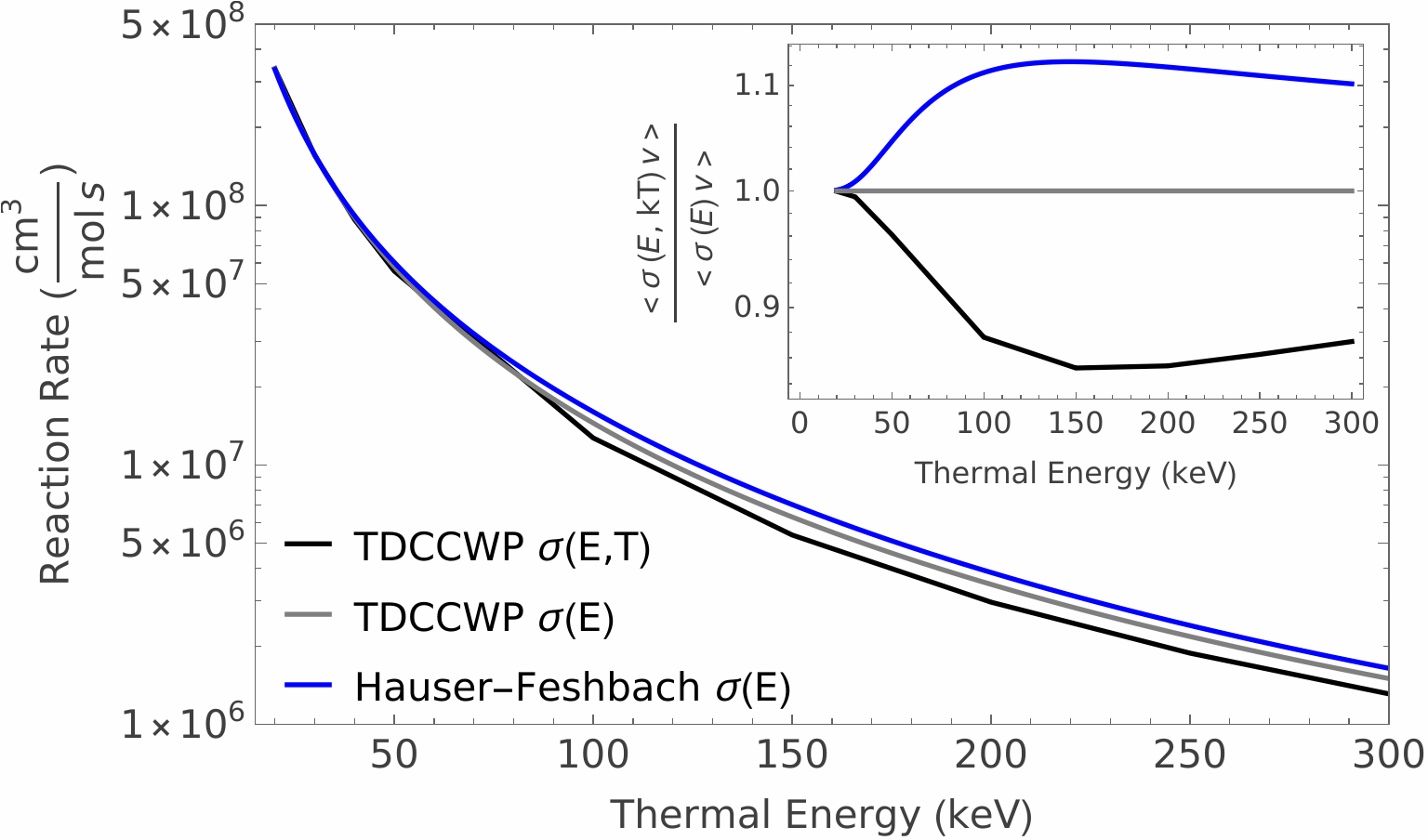}
    \caption{Reaction rates with different types of cross sections from TDCCWP and Hauser-Feshbach style calculations are shown with varied temperature, referred to as thermal energy here, from it being in the form of $kT$.  For TDCCWP there is an agreement in reaction rate for the thermal energies below the first excited state and a decrease in reaction rate for thermal energies far above this.  For $kT=250$ keV, there is a 15$\%$ reduction in reaction rate, which is directly related to the reduction in capture probability with an increase in temperature.  The opposite is seen for the Hauser-Feshbach style calculations, where an increase in 16$\%$ is seen for the same thermal energy.  An inset graph shows the fraction between reaction rates, $\frac{<\sigma(E,T)v>.}{<\sigma(E)v>}$.  The calculation uses a range of incident energies from $E_0=10-1500$ keV.  }
    \label{fig:reactionrate}
\end{figure}

In Fig. \ref{fig:reactionrate}, two distinct regions of the temperature range are present.  The first region shows an agreement between the two TDCCWP with and without temperature dependence for thermal energies below $100$ keV.  The second is where the TDCCWP calculation with thermal effects shows a decrease in reaction rate compared to that of the temperature-independent case.  For low thermal energies, specifically those present in the slow-neutron capture regime, s-process, $kT<30$ keV, an agreement of reaction rates is present.  This may explain why the slow neutron capture regime is dominant for this isotope.  Since the reaction rate only decreases for temperatures much larger than generally present in the s-process, $kT>100$ keV, this could hint at why the s-process dominates the reaction mechanism in the osmium isotopes \cite{humayun2007s}.  This $15\%$ reduction in reaction rates for $kT=250$ keV could highlight the reason why the synthesis of $^{188}$Os is generally not found in r-process environments, but is found in s-process environments at much lower temperatures.  The first step in this process would be to look at isotopes that are dominated by the r-process to show the validity of this statement. 



\subsection{TDCCWP vs Hauser-Feshbach style calculations}

\subsubsection{Cross Sections}
\begin{figure}[t]
    \centering
    \includegraphics[width=\linewidth]{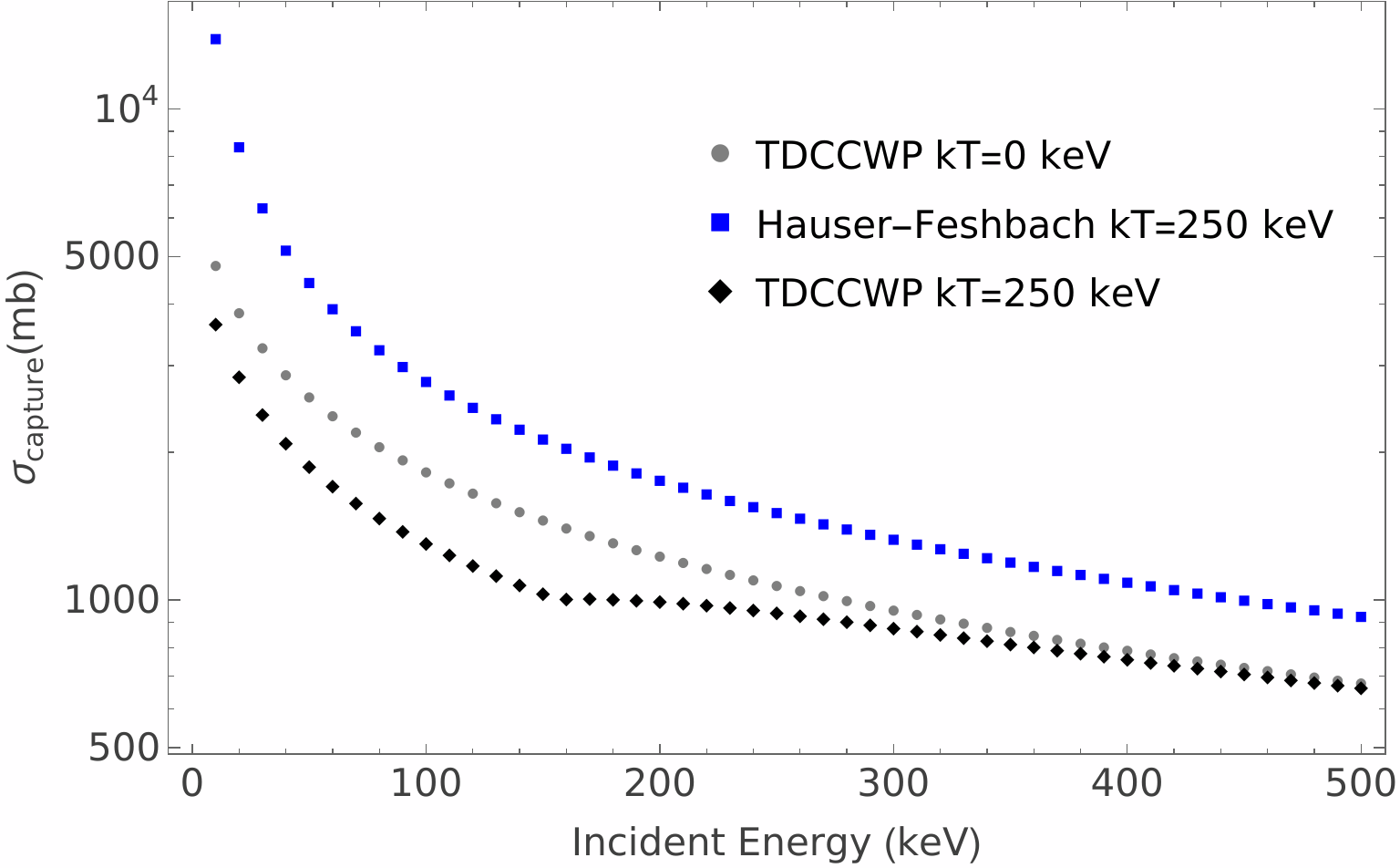}
    \caption{Temperature-dependent capture cross sections for a $l=0$ neutron on the $^{188}$Os target as a function of neutron energy using Hauser-Feshbach style and TDCCWP calculations.  A thermal energy of 250 keV is used in both implementations of thermal effects.  }
    \label{fig:HFTDCCWP}
\end{figure}

To compare with the already existing treatment of thermal effects \cite{mosconi2010neutronI}, a Hauser-Feschbach style implementation of thermal effects on cross sections is applied to the TDCCWP method.  This is done by using a temperature-independent TDCCWP calculation for the neutron capture cross section in a specific excited state of $^{188}$Os and then applying Boltzmann factors in the same way as in Eq. (\ref{HFrates}) for reaction rates. This final state implementation, at the cross-section level, neglects the dynamical coupling among $^{188}$Os excitations, and is compared to the initial state implementation, at the wave-packet level, present in TDCCWP.

This comparison is shown in Fig. \ref{fig:HFTDCCWP} where a reduction in cross section is present when thermal effects are included at the wave-packet level, while an enhancement of cross sections is present when thermal effects are treated at the cross-sectional level.  This hints at the interactions between the ground and excited states having a greater involvement in neutron capture reactions to reduce this enhancement.  An increase in cross sections in the Hauser-Feshbach style calculation is also present for large incident energies, whereas in the TDCCWP model, there is an agreement between the temperature-independent calculation and the temperature-dependent one.  The TDCCWP result also displays the inclusion of the kinematically closed state for incident energies up to the first excited state, 155 keV.

\subsubsection{Reaction Rates}

Previous work has used the Maxwell-Boltzmann velocity distribution to apply thermal effects directly in the calculation of reaction rates separately for the decays of individual excited states \cite{mosconi2010neutronI,fujii2010neutron}.  This Hauser-Feshbach style calculation is done by running a single-channel calculation of TDCCWP for the ground and excited states separately over a range of incident energies from 10-1500 keV.  These energies are shifted by $E=E_0-\epsilon$, so the effective incident energies for the excited states are shifted by the excited state energy, $\epsilon$, to have effective energies in the range 10-1500 keV.  Then each of these components of the cross section are independently used to create reaction rates for each internal energy level.  These reaction rates are then combined to create the total temperature dependent reaction rate in the Hauser-Feshbach style calculation, as explained below.


Previous studies of stellar reaction rates \cite{rauscher2022stellar} were carried out taking the sum of integrals using Boltzmann factors $p_n$:
\begin{eqnarray}
    \label{reactionrate}
    &\langle\sigma v\rangle=\sum_n p_n(\epsilon_n)\langle\sigma v\rangle_n  \label{HFrates} \\
   & = \sqrt{\frac{8}{\pi \mu}} \bigg(\frac{1}{kT}\bigg)^{\frac{3}{2}} \sum_n p_n(\epsilon_n) \frac{\int_{E_i}^{E_f}{\sigma_n(E_n)\cdot E_n \cdot P(E_n,T) dE_n}}{\int_{E_i}^{E_f}{E_n \cdot P(E_n,T) dE_n}}. \nonumber
\end{eqnarray}
In this equation, each $\langle\sigma v\rangle_n$ gives an independent calculation of the reaction rate, where $n$ corresponds to each level of internal energy of the target nucleus \cite{rauscher2022stellar}.  Each sum also involves a shift in incident energies that transforms the integral to adjust for the excitation energy of the excited states of the target, $E_n=E_0-\epsilon_n$, where $E_0$ is the incident energy of the neutron. In addition, these Boltzmann factors $p_n$, as described in subsection 2.4, are used to account for the probability that each excited state has to be populated at a certain thermal energy.  This approach neglects the dynamical coupling between the target's internal thermally populated states during the neutron capture process. However, in TDCCWP the thermal dependence is included from the start of the reaction at the level of the initial wave-function in Eq. (4), which has not been explored in neutron capture cross-sections. As can be seen in Fig. \ref{fig:reactionrate}, the Hauser-Feshbach style calculation displays an increase in reaction rate $16\%$ in contrast to the TDCCWP calculation showing a reduction of $15\%$. This discrepancy hints at the importance of including an initial thermal wave-packet to allow for the interaction of the target's ground and excited states due to the nuclear dynamical coupling as shown by TDCCWP.

\subsubsection{Convergence of reaction rates with respect to the limits of the integral in Eqs. (\ref{reactionrates}) and (\ref{HFrates})}

\begin{figure}[h]
    \centering
    \includegraphics[width=\linewidth]{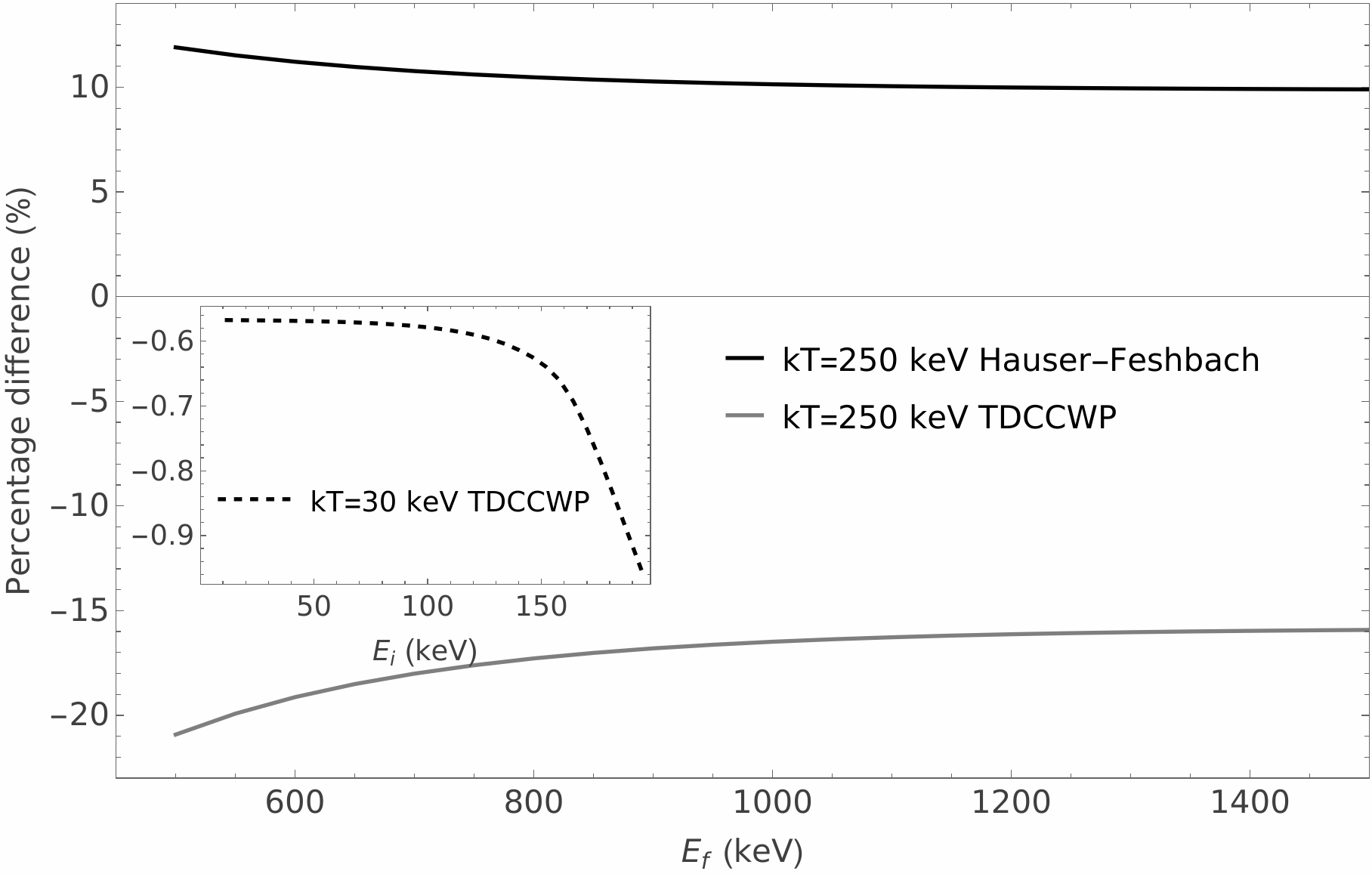}
    \caption{Percentage differences of reaction rates, $100\times\frac{<\sigma(E,T)v>-<\sigma(E)v>}{<\sigma(E)v>}$ for $kT=250$ keV, comparing the Hauser-Feshbach style and TDCCWP calculations. As the upper limit of the integral $E_f$ in Eqs. (\ref{reactionrates}) and (\ref{HFrates}) increases with $E_i$ = 10 keV, a plateau of reaction rates is found for both Hauser-Feshbach style and TDCCWP calculations, retaining the increase or decrease in reaction rate shown in Fig. \ref{fig:reactionrate}.  The inset figure shows the convergence of the percentage difference with respect to the lower limit of the integral $E_i$ in Eqs. (\ref{reactionrates}) and (\ref{HFrates}) for $kT=30$ keV with $E_f=200$ keV.  In this calculation, a range of incident energies $E_0=2-200$ keV was used, which covers 99$\%$ of the Boltzmann distribution.}
    \label{fig:reactionrateconv}
\end{figure}

In order to verify the convergence of the reaction rate calculation with respect to the upper limit of the integral $E_f$ in Eqs. (\ref{reactionrates}) and (\ref{reactionrate}), the contribution of higher incident energies can be included. For example, the Boltzmann distribution predicts that incident energies above 500 keV correspond to $24\%$ of the total probability of the Boltzmann distribution at a thermal energy of $kT=250$ keV.  In Fig. \ref{fig:reactionrateconv}, it can be seen that for $E_f > 1500$  keV there is a plateau of reaction rate results. This convergence still retains the reduction in reaction rate for the TDCCWP calculation and an increase in reaction rate in the Hauser-Feshbach style calculation, as was seen in the previous section.  Therefore, the selected incident energy range is 10-1500 keV for the calculation of reaction rates both for the Hauser-Feshbach and TDCCWP calculations in Fig. \ref{fig:reactionrate}. For a thermal energy of $kT=250$ keV, the probabilities corresponding to incident energies below 10 keV are less than 1$\%$.  Therefore, the opposite convergence is not shown for this thermal energy. There is a similar plateau of percentage differences in reaction rates for decreasing values of the lower limit of the integral $E_i$ in Eqs. (\ref{reactionrates}) and (\ref{HFrates}), where there is a convergence once $E_i$ reaches 10 keV. This is shown in the inset of Fig. \ref{fig:reactionrateconv} for the thermal energy $kT$ = 30 keV and an incident energy range of $E_0=2-200$ keV. In this calculation, an initial wave-packet with spatial width of $\sigma_0=230$ fm was used, which has an energy variance of $\Delta E=0.19$ keV. For thermal energies in the interval $kT=30-300$ keV, the range of incident energies $E_0=10-1500$ keV corresponds to $>98\%$ of the Boltzmann distribution.   

\section{Conclusion}

This paper presents the TDCCWP model to show the importance of thermal effects in reaction cross-sections in the $^{188}$Os+n reaction as a test case.  Transmission coefficients were generated to calculate cross-sections and reaction rates, including thermal effects at the initialisation step.  Firstly, the comparison between the TDCCWP and CCDM methods was shown to agree for high incident energies and shows a decrease in the capture probability in TDCCWP for low incident energies when thermal effects are included.  Subsequently, temperature-dependent cross sections were shown to exhibit a decrease in cross section for an increase in temperature, contrary to applying thermal effects to cross sections, as commonly done in the literature.  This change was highlighted by a decrease in reaction rates at high thermal energies, up to a $15\%$ reduction at $kT=250$ keV, which may hint at an importance in including thermal effects in this way for r-process reactions.  The main physical reason for the difference caused by thermal effects on reaction rates in the TDCCWP and Hauser-Feshbach style calculations is the crucial role of the dynamical nuclear coupling between the thermally populated target's states, which is neglected in the Hauser-Feshbach style approach, but creates a dominant neutron capture pathway with an increased neutron speed and thus reduces the neutron capture cross section.      

Further work will address thermal effects in neutron capture reactions involving osmium and rhenium isotopes.  This would give a better understanding of the contribution of thermal effects in the neutron capture cross sections throughout the Re-Os chain.  In addition, this could elaborate on the temperature-dependent contribution of the abundance of $^{187}$Os from neutron capture reactions.  This would work in contrast to the abundance of $^{187}$Os from the $\beta_-$ decay from $^{187}$Re, ultimately giving another estimate for stellar enhancement factors and for the age of the universe using the Re-Os clock \cite{vertebny1969slow}.  Subsequently, other isotopes could be explored, such as $^{99}$Zr, which is specifically applicable as an unstable byproduct in nuclear reactors.  Finally, californium, relevant to the r-process, can be studied to highlight the change in cross sections and reaction rates at high temperatures present in these reactions, where the results differ the most.  These extensions are valuable both for low-energy neutron physics and for exploring the resonance regions of these isotopes.
\section{Acknowledgments}
This was supported by the UK Science and Technology Facilities Council (STFC) under Grants No. ST/Y509619/1, ST/V001108/1 and ST/Y000358/1.
\section{Data Availability}
The data supporting the findings of this article are not publicly available.  The data is available from the authors upon reasonable request.
\printbibliography
\end{document}